\documentclass[fleqn,usenatbib]{mnras}

\usepackage{newtxtext,newtxmath}
\usepackage[T1]{fontenc}

\DeclareRobustCommand{\VAN}[3]{#2}
\let\VANthebibliography\thebibliography
\def\thebibliography{\DeclareRobustCommand{\VAN}[3]{##3}\VANthebibliography}
\newcommand{\sn}{{\rm S/N}} 
\usepackage{color}
\usepackage{ulem}
\usepackage{graphicx}
\usepackage{amsmath}
\usepackage{subcaption}
\usepackage{changes}

\title[MW FRBs with BURSTT]{Detection Rate of Fast Radio Bursts in the Milky Way with BURSTT}

\author[Ling et al.]{Decmend Fang-Jie Ling,$^{1}$
Tetsuya Hashimoto,$^{2}$
Shotaro Yamasaki,$^{2}$
Tomotsugu Goto,$^{3}$
Seong Jin Kim,$^{3}$
\newauthor{Simon C.-C. Ho,$^{3}$
Tiger Y.-Y. Hsiao,$^{3}$
and Yi Hang Valerie Wong$^{3}$}
\\
$^{1}$Department of Physics, National Tsing Hua University, No. 101, Section 2, Kuang-Fu Road, Hsinchu City 30013, Taiwan (R.O.C.)\\
$^{2}$Department of Physics, National Chung Hsing University, No. 145, Xingda Rd., South Dist., Taichung, 40227, Taiwan (R.O.C.)\\
$^{3}$Institute of Astronomy, National Tsing Hua University, 101, Section 2. Kuang-Fu Road, Hsinchu, 30013, Taiwan (R.O.C.)\\
}

\date{Accepted 2022 October 17. Received 2022 October 17; in original form 2022 June 13}

\pubyear{2022}

\begin{document}
\label{firstpage}
\pagerange{\pageref{firstpage}--\pageref{lastpage}}
\maketitle

\begin{abstract}
Fast radio bursts (FRBs) are intense bursts of radio emission with durations of milliseconds. Although researchers have found them happening frequently all over the sky, they are still in the dark to understand what causes the phenomena because the existing radio observatories have encountered certain challenges during the discovery of FRB progenitors. The construction of Bustling Universe Radio Survey Telescope in Taiwan (BURSTT) is being proposed to solve these challenges. We simulate mock Galactic FRB-like events by applying a range of spatial distributions, pulse widths and luminosity functions. The effect of turbulent Interstellar Medium (ISM) on the detectability of FRB-like events within the Milky Way plane is considered to estimate the dispersion measure and pulse scattering of mock events. We evaluate the fraction of FRB-like events in the Milky Way that are detectable by BURSTT and compare the result with those by Survey for Transient Astronomical Radio Emission 2 (STARE2) and Galactic Radio Explorer (GReX). We find that BURSTT could increase the detection rate by more than two orders of magnitude compared with STARE2 and GReX, depending on the slope of luminosity function of the events. We also investigate the influence of the specifications of BURSTT on its detection improvement. This leads to the fact that greatly higher sensitivity and improved coverage of the Milky Way plane have significant effects on the detection improvement of BURSTT. We find that the upgrade version of BURSTT, BURSTT-2048 could increase the detection rate of faint Galactic FRB-like events by a factor of 3.

\end{abstract}


\section{Introduction}

Fast Radio Bursts (FRBs) are transient radio pulses with typical duration of a few milliseconds, of which progenitors are unknown \citep{Lorimer2007, Thornton2013}. FRBs are a common phenomenon since their appearance rate is approximately 1000 per day \citep[e.g.,][]{Bhandari2018} in the sky. FRBs have characteristic features in that their pulse arrival time delays as a function of frequency due to the propagation through an intervening ionised medium. This delay is described by an observed quantity, so called dispersion measure (DM), defined as a free electron number density integrated along a line of site. FRBs are broadly divided into repeating and non-repeating classes depending on whether they repeat multiple times or not, while most FRB sources are classified as non-repeating FRBs. However, it is so far unclear whether apparently non-repeating FRBs are truly non-repeaters \citep[e.g.,][]{Chen2022, Hashimoto2022,Ai2021}, which is essential information to understand both their progenitors and emission mechanisms.

During the progress in understanding the origin of FRBs, most of the observatories have encountered a lot of challenges. First of all, the current survey telescopes lack a localization capability. Since FRBs disappear in short timescales, it causes the accurate positions of FRB progenitors to be hardly measured without immediate localization. To date, despite that more than ~600 FRBs have been discovered \citep[e.g.,][]{Petroff2016,CHIMEcat2021}, there is only one confirmed FRB progenitor, which was identified as a Galactic magnetar, SGR 1935+2154. It was detected by both Survey for Transient Astronomical Radio Emission 2 (STARE2) and Canadian Hydrogen Intensity Mapping Experiment (CHIME) \citep{Bochenek2020b, Scholz2020}. However, the insufficient localization accuracy of current survey telescopes (for instance, CHIME has localization uncertainties of arcminutes for most bursts \citep{CHIME/FRBCollaboration2018}) has been hampering identification of the other FRB progenitors even though these survey observations have already operated for a few years. Therefore, localization of FRBs becomes the most important part to discover the origin of these events and the proposal of a telescope with higher localization accuracy is essential.

Furthermore, the observation conducted by most FRB survey telescopes could not cover a wide coverage of the sky continuously with a very high cadence. FRB’s repetitive nature cannot be understood because of the limited observational time used by the current survey telescopes. Identifying whether these events are repeaters or non-repeaters is essential to reveal their mysterious origins since repeating FRBs may be generated by the progenitors with repeating activities, such as pulsars and magnetars, while the origins of non-repeating FRBs could be one-off events, such as compact merger systems \citep{Platts2019}. However, without such long-term monitoring observations, a significant fraction of repeating bursts may be missed. Some non-repeating FRBs may not be one-off events, but repeat events can be detected from them in future follow-up observations or repeat events might have been missed in the previous observations. Thus, such FRBs could be misclassified as non-repeating FRBs. 

The discovery of a short and intense galactic FRB from SGR 1935+2154 by both STARE2 and CHIME was exactly what researchers had been missing. Exploring the nearby Universe where we can maximize the chance of detecting any multi-wavelength counterparts of FRBs becomes an important way to reveal the origin of FRBs. Nevertheless, the optimal design for simultaneous multi-wavelength observations is not employed in current telescopes which leads to the difficulty of detecting multi-wavelength counterparts of FRBs discovered to date. Multi-wavelength counterpart search has so far been limited by distance and most of the detected FRBs are located too far away.

To solve the challenges in revealing the physical origins of FRBs, the telescope dedicated to detecting FRB-like events in the nearby Universe is proposed. Bustling Universe Radio Survey Telescope in Taiwan (BURSTT) is a radio software telescope with unique fisheye and it consists of a main 256 antenna array in Taiwan with smaller outriggers in Hawaii and at other locations in Taiwan \citep{Lin2022}. The BURSTT project plans to upgrade its main antenna array up to $\sim$2,000 antennas in the future. BURSTT has Log-Periodic Dipole Array (LPDA) antennas as receivers with approximately 150$\times$150 cm for each antenna. Its specification includes a 1.52 steradians field of view (FoV) and 400 MHz of the spectral bandwidth centered at 0.6 GHz. Due to its unique fisheye design and extremely wide FoV, BURSTT will observe 25 times more of the sky than CHIME to prevent missing FRB-like events easily. The outriggers of BURSTT will provide approximately 1 arcsecond localization for the host galaxy via very-long baseline interferometry (VLBI), which is more accurate than the current telescopes. Moreover, BURSTT will conduct long term monitoring observations to prevent missing any repeating FRBs, which will resolve the missing repeating FRB problem \citep[e.g.,][]{Ai2021}.

In this work, we investigate what fraction of simulated FRB-like events in the Milky Way are detectable with a BURSTT-liked instrument, considering its sky coverage, single pulse sensitivity, instrumental time resolution and the DM smearing due to the frequency channelisation. This paper is organized as follows. We describe how we model FRB-like events in the Milky Way, including the spatial distributions, total pulse widths and pulse luminosities of these events in Section \ref{sec:Modelling FRB-like events in the Milky Way}. The effects of the Interstellar Medium (ISM) on the detection of FRB-like events in the Milky Way are considered when setting up distributions of FRB-like event sources in the Milky Way since pulses travelling through the ISM are scattered as a function of DM \citep{Bhat2004}. We then simulate the recently proposed BURSTT experiment to investigate how it probes the Milky Way for bursts and evaluate the performance of BURSTT \citep{Lin2022} over those of STARE2 \citep{Bochenek2020a} and Galactic Radio Explorer (GReX) \citep{Connor2021} in Section \ref{sec:Simulation of BURSTT survey}. STARE2 and GReX are chosen since the comparison of performance between these two surveys has been proposed in the previous paper \citep{Gohar2022}. In Section \ref{sec:Factors of specification of BURSTT in detection rate improvement}, we discuss which of the specifications of BURSTT most significantly affects its FRB-like events search efficiency. The performance of the recently proposed BURSTT with 2048 antenna (BURSTT-2048) is evaluated in Section \ref{sec:BURSTT with 2048 antennas}. We summarise and conclude in Section \ref{sec:Conclusion}.

\section{Modelling FRB-like events in the Milky Way}
\label{sec:Modelling FRB-like events in the Milky Way}

In this section, we model populations of FRB-like events in the Milky Way by applying a range of FRB-like event spatial distributions, total pulse widths and pulse luminosities. Following \citet{Gohar2022}, we use the publically available code, \texttt{MilkyWay-FRBs} \citep{Gohar2022}, to simulate the Galactic FRB-like events in this work.

\subsection{Spatial distributions}

So far there is a lack of the discovery of FRB-like event sources in our local galaxy and their spatial distributions are still unknown, but we can still simulate them according to the related phenomena that happened in the Universe.

The observations of an FRB-like event from a magnetar in the Milky Way (SGR 1935+2154) by the CHIME and STARE2 telescopes \citep{Bochenek2020b, Scholz2020} imply that FRB sources are associated with star-forming regions and ionised ISM. Besides that, an FRB that is associated with a globular cluster in a nearby galaxy has been detected \citep{Bhardwaj2021}, and it illustrates that FRB sources also appear in old stellar populations. The old populations are also suggested from the redshift evolution of number densities of non-repeating FRBs \citep[e.g.,][]{Hashimoto2022,Zhang2022}. By considering these two cases, we assume that FRB-like event sources are distributed in accordance with very young stars or general stellar populations. We use a publicly available code, \texttt{MilkyWay-FRBs} \citep{Gohar2022},  to construct the following models to generate the mock FRB-like events in Milky Way:

\begin{itemize}
\item Model I: the FRB-like events are radially distributed in an exponential disk, that is the surface density exponentially decrease with increasing distance from the Galactic center with a scale-length $R_{\rm h}$:
\begin{equation} \label{eq:01}
\rho \propto e^{-R/R_{\rm h}}e^{-|z|/z_{\rm h}}\ ,
\end{equation}
where $R = (X+Y)^{1/2}$ is the projected distance of FRB-like event on the Galactic plane measured from the Galactic center, $R_{\rm h}$ is the scale length of the disk, exponential $z$ is the vertical distribution of FRB-like events and $z_h$ is the exponential scale height. We adopt $R_{\rm h}  = 4$ kpc as representative of the old stellar disk \citep{Lewis1989}.
The ionized ISM is modelled using one of the publicly available codes, NE2001 \citep{Cordes2002}.
    
\item Model II: the FRB-like events are located along the ionised ISM in spiral arms: 
\begin{equation} \label{eq:02}
\rho \propto n_e(R) e^{-|z|/z_{\rm h}}\ ,
\end{equation}
where $n_e(R)$ is the electron density as a function of Galactocentric radius, which is modelled using another publicly available code, YMW16 \citep{Yao2017} along the line-of-sight.
\end{itemize}

By adopting these models, we create 883073 and 879166 FRB-like event source 2D (X, Y) positions (in Galactocentric Coordinates) in the disk with a surface density in proportion to the stellar density distribution (Model I) and the electrons in the ISM (Model II) respectively. The vertical distribution of the events is assumed to have exponential scale height of $z_{\rm h}=50$ pc for both Model I and II based on magnetar observations \citep[e.g.,][]{Kaspi2017}. 

\subsection{Total pulse widths}

Here we briefly describe basic assumptions made in the code, \texttt{MilkyWay-FRBs} \citep{Gohar2022} that we utilise in this work. To calculate the total pulse width $\tau_{\rm pulse}$ of each simulated FRB-like event, the following equation is applied:
\begin{equation} \label{eq:03}
\tau_{\rm pulse} = (t_{\rm scat}^2 + t_{\rm  int}^2 + t_{\rm DM}^2)^{1/2}\ ,
\end{equation}
where $t_{\rm scat}$ is the pulse broadening time due to scattering, $t_{\rm int}$ is the intrinsic width of each FRB-like event and $t_{\rm DM}$ is the dispersion measure (DM) smearing due to the finite channel widths of an operating telescope.

The pulse scattering effect is included when computing the total pulse width of the events because the detectability of pulse is strongly affected by the properties of ISM since pulses travelling through the ISM are scattered as a function of DM \citep{Bhat2004}. The DM of each FRB-like event due to the ISM models (YMW16 and NE2001) is calculated with:
\begin{equation} \label{eq:05}
{\rm DM} = \int_{0}^{d} n_e(l)dl \ ,
\end{equation}
where $d$ is the distance from the FRB-like event to the observer (telescope) and $n_e$ is the electron density along the line-of-sight.
The Bhat relation is used to obtain the pulse broadening time:
\begin{equation} \label{eq:04}
\log(t_{\rm scat,-3}) = -6.46 + 0.154x + 1.07x^2 - 3.9\log(\nu_9)\ ,
\end{equation}
where $t_{\rm scat,-3}=t_{\rm scat}/10^{-3}$ is the pulse broadening time in ms, $\nu_9=\nu/10^9$ is the observing frequency of telescope in GHz and $x \equiv \log[{\rm DM}/({\rm pc\, cm^{-3}})]$ relates to the DM for the pulse. 
The DM smearing (measured in µs) due to the finite channel widths of a telescope is given by:
\begin{equation} \label{eq:06}
t_{\rm DM} = 8.3 \, {\rm DM} \,\Delta\nu_6 \, \nu_9^{-3}\,{\rm \mu s},
\end{equation}
where $\Delta\nu_6=\Delta\nu/10^6$ is the width of the frequency channels in MHz.

An important notice is that there is an observational scatter around the pulse width-DM relation (equation \ref{eq:04}) with approximately a factor of 10 for a lognormal distribution of FRBs. Therefore, A lognormal scatter of a factor of 10 around the Bhat relation is applied to each simulated FRB-like event in our modelling. This procedure did not change the detectability rates of FRB-like events very much overall, but it affects the range of distances from the Sun and DMs for detectable events \citep{Gohar2022}.

In our modelling, we set a pulse width of 100 ms as a threshold below which FRB-like events can be searched. In typical FRB research, it is a maximum value of pulse width that can be searched before the observing systems become overwhelmed by the false positives due to Radio Frequency Interference (RFI).

\subsection{Signal to noise ratio}

To simulate FRB-like event luminosities, a luminosity function (LF) is assumed and a signal-noise ratio (S/N) is assigned to each simulated FRB-like event. The LF of FRB-like events is modeled as a power-law with a lower energy cut-off, $E_0$:
\begin{equation} \label{eq:07}
\frac{dN}{dE} \propto \left({E \over E_0}\right)^\alpha\ ,
\end{equation}
where $E$ is the intrinsic energy of the bursts, $E_0$ is a lower energy cut-off to the distribution, and $\alpha$ is the power-law index.

For the events that survive the maximum width cut, which means that the pulse width does not exceed 100 ms, each FRB-like event is allocated with an S/N in the simulation by using the radiometer equation below:
\begin{equation} \label{eq:08}
\sn = {S_{\rm peak} \over S_{\rm lim}}(BW \times N_{\rm pol} \times \tau)^{1/2}\ ,
\end{equation}
where $S_{\rm peak} = E⁄(4 \pi d^2 \tau \nu)$ is the peak flux density of the burst with $d$ being the distance from the FRB-like event to the observer and $\tau$ being the pulse width of an FRB-like event, $S_{\rm lim}$ is the system equivalent flux density (SEFD) of the operating telescope, $BW$ is the system bandwidth, and $N_{\rm pol}$ is the number of polarization (we adopt $N_{\rm pol} = 2$).

Among the mock FRB-like event sources, we remove the events with $\sn < 10$ from our sample in Section 3.2, since $\sn = 10$ is the typical detection threshold adopted by past and ongoing radio transient surveys.

\section{Simulation of BURSTT survey}
\label{sec:Simulation of BURSTT survey}

BURSTT is a proposed instrument tailored for detecting FRB-like events with accurate localization and high cadence \citep{Lin2022}. BURSTT will observe 25 times more of the sky than CHIME, because of its unique fisheye design and FoV of 1.52 steradians (see Table \ref{table:1}). At the same time, BURSTT is an interferometer with baselines of over 100 km and has the great sensitivity, allowing us to discover a large sample of bright FRB-like events with localization, including those located close to the Earth.

In this section, we apply the system parameters of BURSTT to our models in order to analyse the observational properties of the FRB-like events in the Milky Way that can be detected via BURSTT. The relative number of FRB-like events to be detected with BURSTT is compared with those of GReX and STARE2.

\subsection{Simulation of mock FRB-like events in Milky Way}
\label{subsec:Simulation of mock FRB-like events in Milky Way}

We adopt the sensitivity, observing frequency and bandwidth of BURSTT to equations \ref{eq:04}, \ref{eq:06} and \ref{eq:08}, followed by the run of simulations. We apply a lower energy cut-off of $10^{27}$ erg for the simulation. This cut-off energy is much lower than that of the intense radio burst from the Galactic magnetar SGR 1935+2154:  approximately $10^{34}$--$10^{35}$ erg according to its fluence and the estimated distance to SGR 1935+2154 \citep{CHIME/FRBCollaboration2020}. Therefore, it sufficiently samples FRB-like events at the low end of the energy scale while not over-producing FRB-like events that are too dim to be detected \citep{Gohar2022}. We apply a range of power-law slopes of $\alpha= -0.3, -0.5$ and $-1.0$ in our simulation. We set a power-law slope of $\alpha = -0.5$ as a fiducial value in our modelling. The results of a run of simulation of FRB-like events for Model II with $\alpha= -0.3$ and $-1.0$ are shown in Appendix \ref{sec:Testing difference slopes of luminosity functions} (the effect of different slope values on the detection rates is discussed in Section \ref{sec:Comparison of detection rates among STARE2, GReX and BURSTT}).

\begin{figure*}
    \includegraphics[width=17cm]{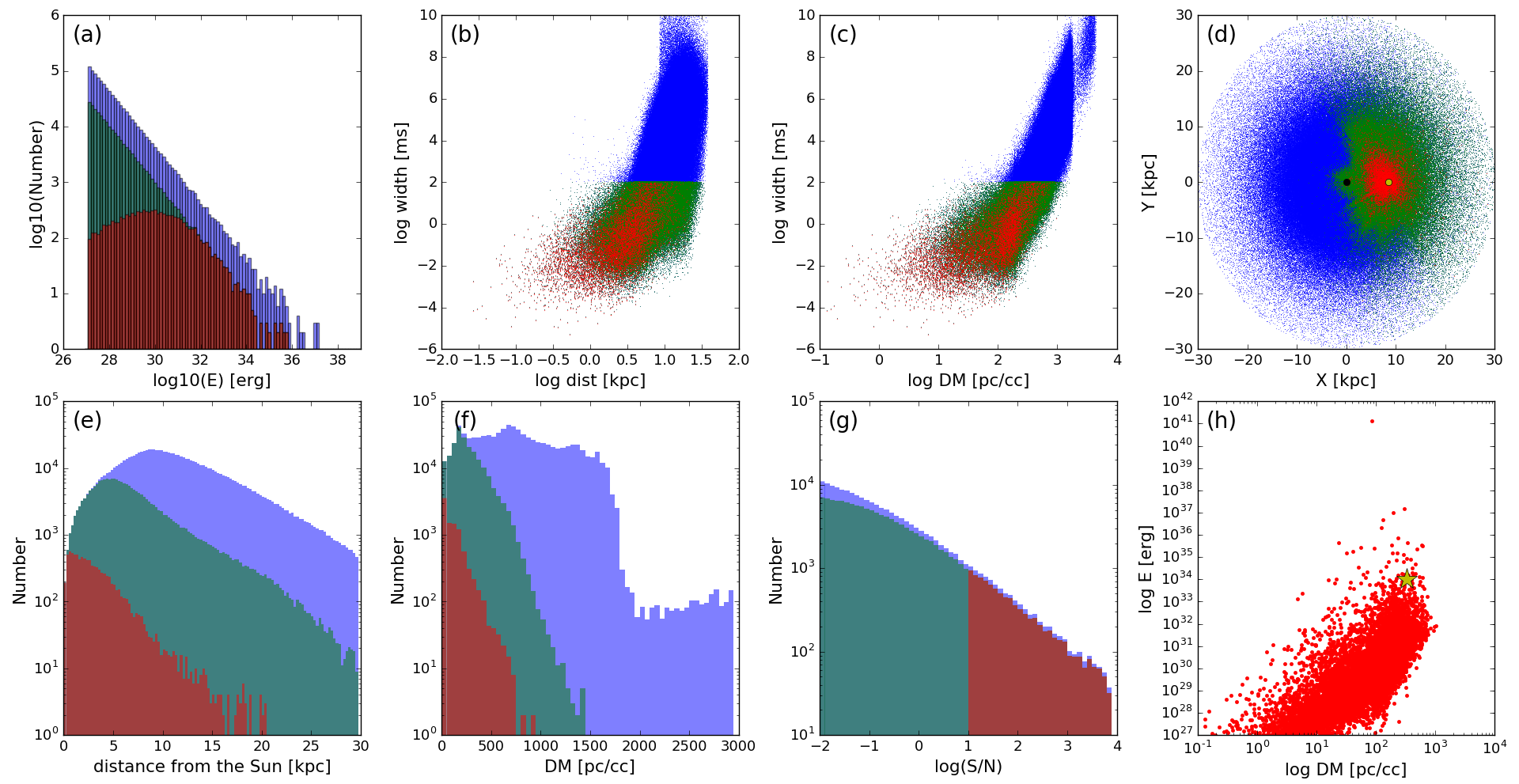}
    \caption{Simulated FRB-like events in the Milky Way for Model I, with power-law index $\alpha = -0.5$ and lower energy cut-off $E_0 = 10^{27}$ erg. In all panels, blue represents all FRB-like events simulated, green represents those that pass the width cut $\tau < 100$ ms and red represents those that pass the detection threshold $\sn > 10$ and $\tau < 100$ ms. The yellow and black dots in panel (d) represent the Sun and the Galactic center, respectively. The yellow star symbol in panel (h) represents the Galactic FRB event from SGR 1935+2154 detected in April 2021. The system parameters of BURSTT are assumed during the simulation \citep{Lin2022}.}
    \label{fig:01}
\end{figure*}

\begin{figure*}
    \includegraphics[width=17cm]{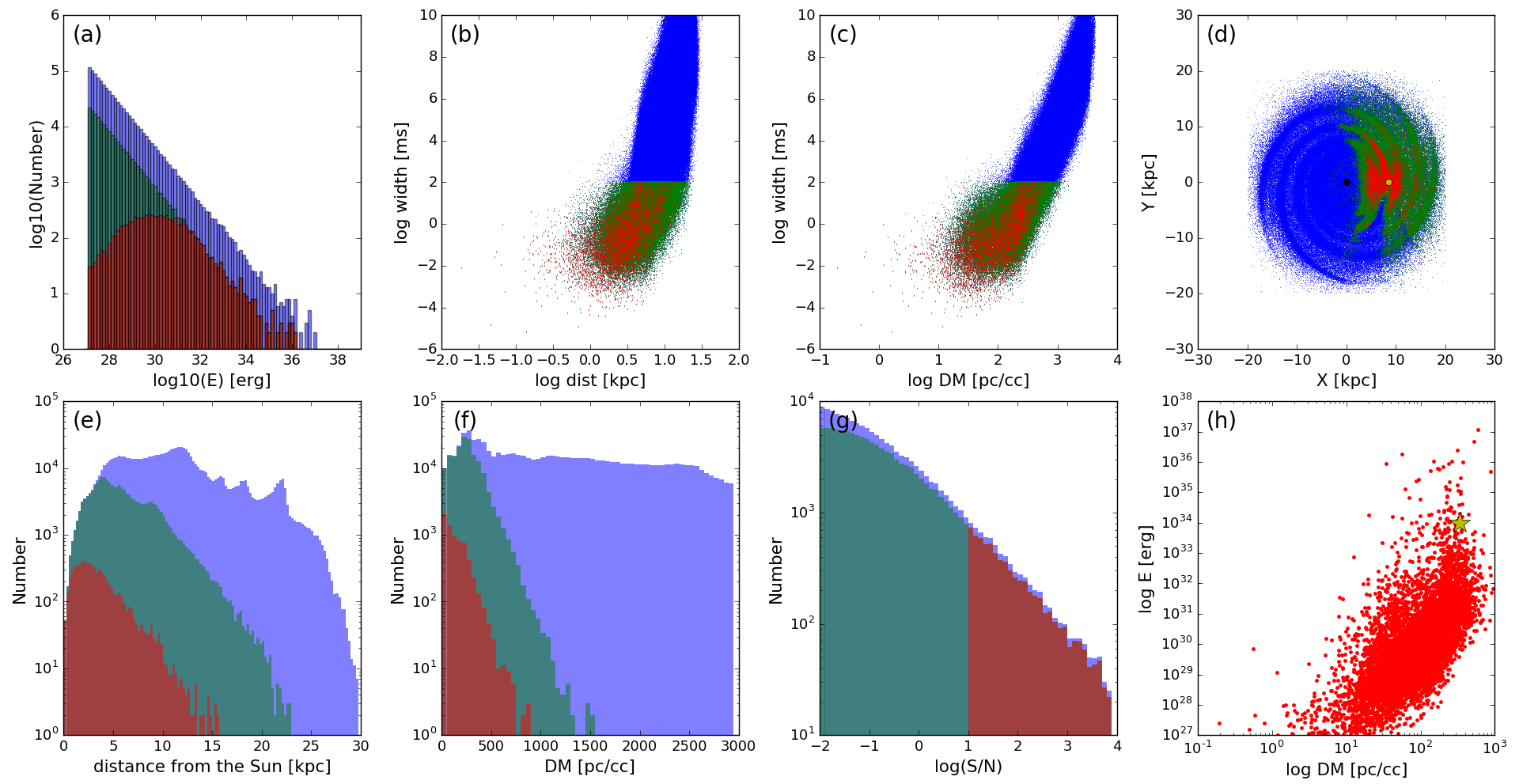}
    \caption{Simulated FRB-like events in the Milky Way for Model II, with power-law index $\alpha = -0.5$ and lower energy cut-off $E_0 = 10^{27}$ erg. All panels are the same as Figure \ref{fig:01}.}
    \label{fig:02}
\end{figure*}

We examine the effect of ISM on the detection of FRB-like events by utilizing two publicly available ISM models, NE2001 and YMW16. The results of a run of simulation of FRB-like events for Models I and II with a power-law of $\alpha = -0.5$ are shown in Figures \ref{fig:01} and \ref{fig:02} and respectively. In all panels, blue represents all simulated FRB-like events, green represents those that pass the width cut $\tau < 100$ ms and red represents those that pass both the detection threshold $\sn > 10$ and $\tau < 100$ ms. In both figures, panel (a) shows the power-law energy distribution of the FRB-like events. Panel (b) shows the pulse width of the FRB-like events as a function of distance from the Sun. Panel (c) shows the pulse widths versus the DM along the line of sight. Panel (d) shows the (X, Y) locations of the FRB-like events in the Milky Way disk. In this panel, the exponential disk model with NE2001 (Model I) is adopted in Figure \ref{fig:01} and the YMW16 model (Model II) in Figure \ref{fig:02}. Panel (e) shows the distribution of distances of FRB-like events from the Sun. Panel (f) shows the DM distribution of the FRB-like events. Panel (g) shows the S/N distribution of the FRB-like events. Panel (h) shows the power-law energy of the FRB-like events as a function of DM.

Based on both Figures \ref{fig:01} and \ref{fig:02}, a typical detectable FRB-like event by BURSTT has an energy of approximately $10^{27}$ to $10^{37}$ erg and a DM in the range of 0 to 900 $\rm{pc\,{cm}^{-3}}$. However, for Model I (Figure \ref{fig:01} panel (e)), the detectable FRBs are located in the range between 0 and 20 kpc from the Sun, while the range of distribution is between 0 and 15 kpc for the Model II (Figure \ref{fig:02} panel (e)). It is noted that in Figure \ref{fig:02} panel (e), there is a spiky structure in the distance distribution of FRB-like events from the Sun because Model II is applied in the simulation and it is akin to the young stellar population which is heavily confined to spiral arms. We note that the absolute numbers of FRB-like events shown in Figures \ref{fig:01} and \ref{fig:02} depend on the total number of simulated FRB-like events. Therefore, the relative number would be meaningful when it is compared with those predicted from the other telescopes including STARE2 and GReX. This point is discussed in the following section.

\begin{table}
    \centering
    \caption{Specifications of STARE2, GReX and BURSTT for search.}
    \begin{tabular}{lccc} 
     \hline
     \hline
     Specification & \multicolumn{3}{|c|}{Value} \\
     \hline
     \hline
     & STARE2 & GReX & BURSTT \\
     \hline
     Area of antenna & $0.2 \times 0.2 $ m$^2$ & $0.2 \times 0.2 $ m$^2$ & $1.5 \times 1.5$ m$^2$ \\ 
     \hline
     SEFD & 19.17 MJy & 3.83 MJy & 5 kJy \\ 
     \hline
     Central frequency & 1.4 GHz & 1.4 GHz & 0.6 GHz \\ 
     \hline
     Bandwidth & 188 MHz & 1700 MHz & 400 MHz \\
     \hline
     Field of view (FoV) & 1.12 str & 1.5 str & 1.52 str \\
     \hline
    \end{tabular}
    \label{table:1}
\end{table}

\subsection{Comparison of detection rates among STARE2, GReX and BURSTT}
\label{sec:Comparison of detection rates among STARE2, GReX and BURSTT}

\begin{figure}
    \begin{subfigure}{10cm}
    \includegraphics[width=9.5cm]{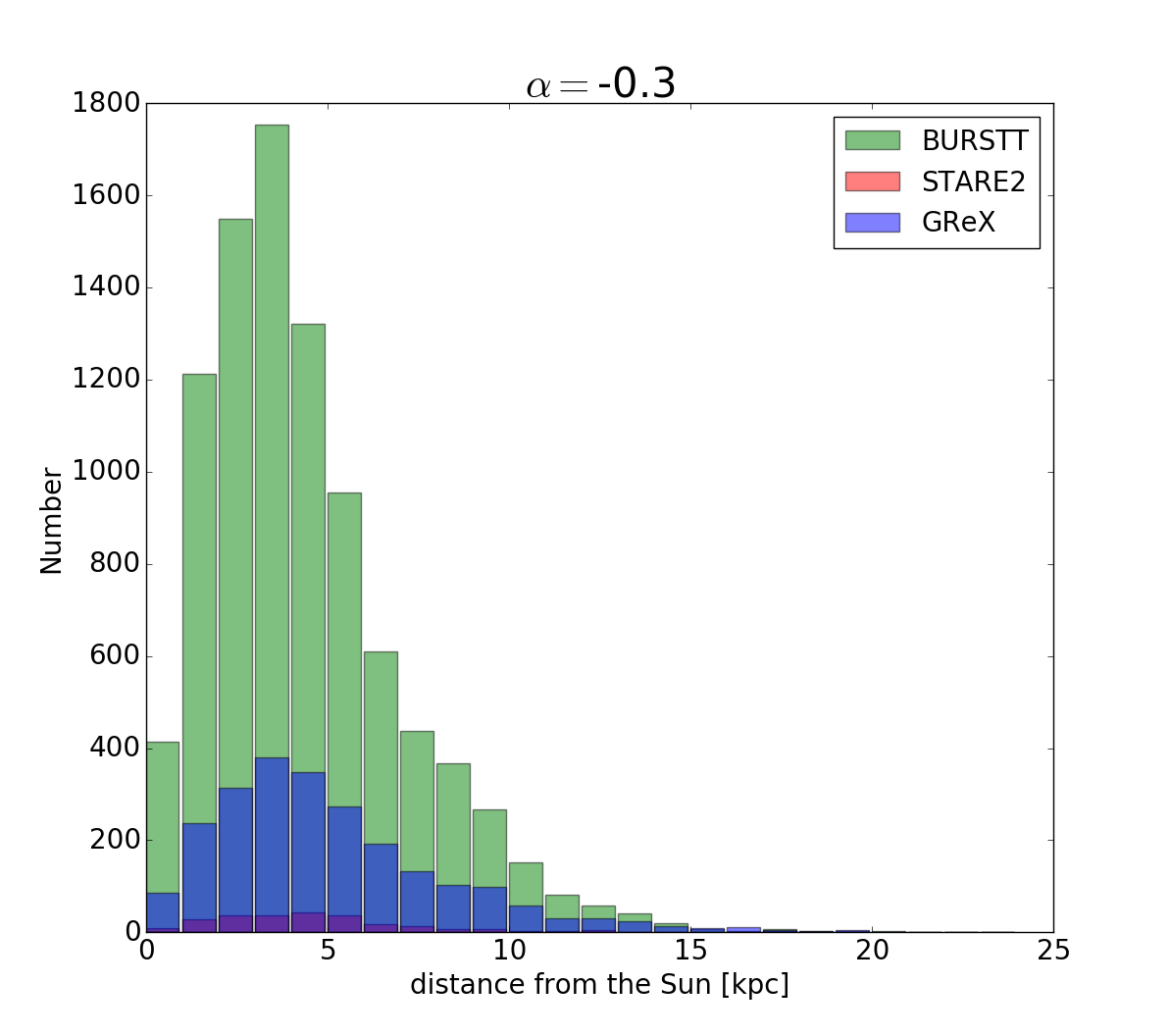}
    \end{subfigure}
    \begin{subfigure}{10cm}
    \includegraphics[width=9.5cm]{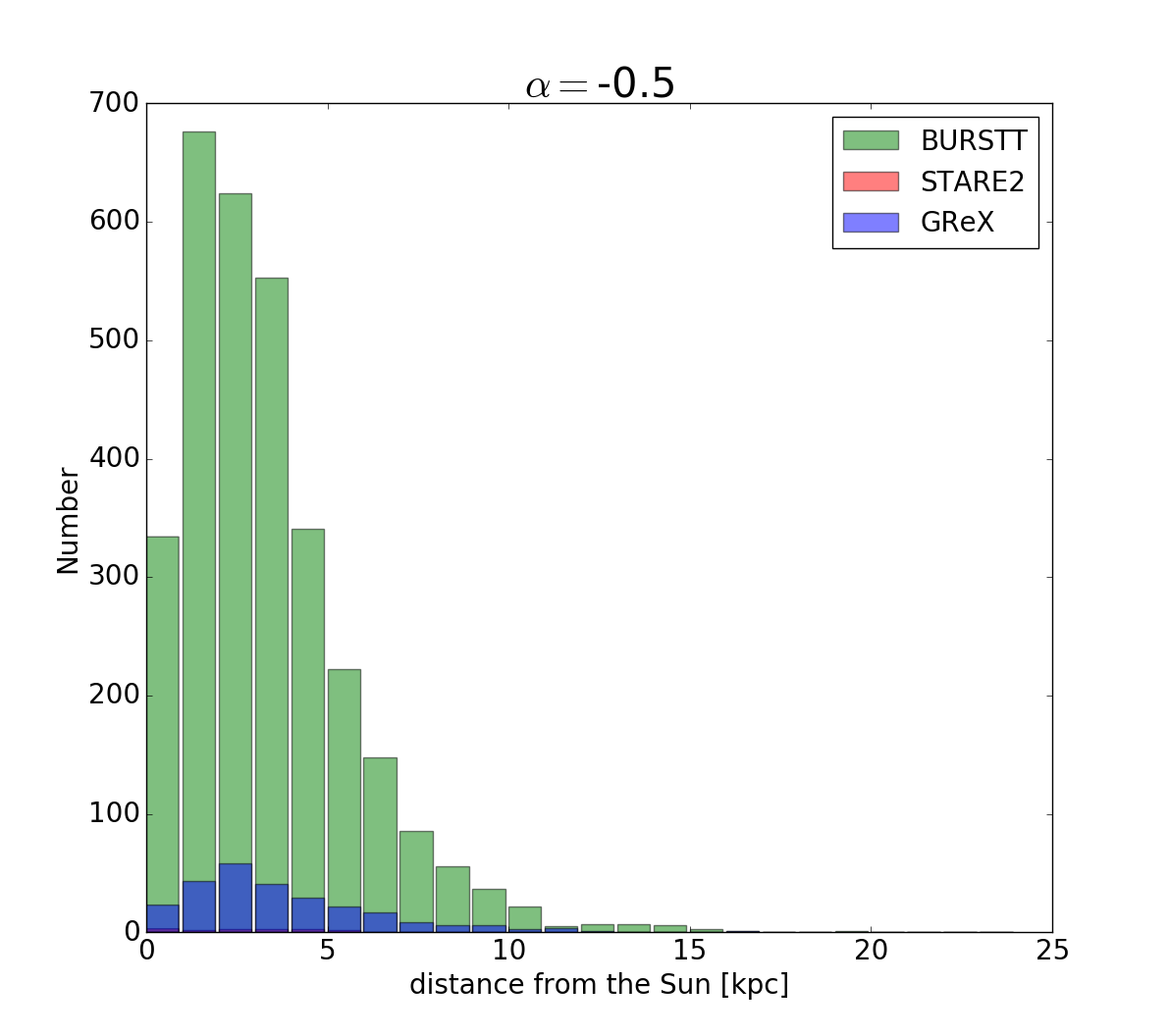}
    \end{subfigure}
    \caption{Upper panel: Comparison of STARE2, GReX and BURSTT detection rates for Model II, $E_0 = 10^{27}$ erg for the LF with $\alpha = -0.3$. The vertical axis represents the number of FRB-like events that pass both the detection threshold ($\sn > 10$) and width cut ($\tau < 100$ ms). Lower panel: Same as the upper panel but with an LF with $\alpha = -0.5$. The detection rate for STARE2 (shown in red color) is much smaller than BURSTT (shown in green color) in this case.}
    \label{fig:03}
\end{figure}

Thus far we have investigated the observational properties of the mock FRB-like events that are detectable via BURSTT. It is scientifically useful to estimate the improvement in detection rates of FRB-like events for BURSTT over the state-of-the-art FRB-search instruments, STARE2 and GReX. Therefore, by using the proposed system parameters of STARE2 \citep{Bochenek2020a}, GReX \citep{Connor2021}, and BURSTT \citep{Lin2022} in equations \ref{eq:04}, \ref{eq:06} and \ref{eq:08}, we run our models for the YMW16 ISM model with lower energy cut-off $E_0 = 10^{27}$ to compare the discovery rates among these three surveys. A range of power-law slopes of $\alpha= -0.3, -0.5$ and $-1.0$ is adopted in the simulation to examine its effect on the detection rates of FRB-like events. Such range of power-law slopes is chosen based on the constraints of the rate of bursts which are obtained from the CHIME/FRB detection from SGR 1935+2154 and the CHIME/FRB non-detection of bursts from nearby star-forming galaxies \citep{CHIME/FRBCollaboration2020}. The results of simulations of FRB-like events for Model II with power-law $\alpha = -0.3$ and $-0.5$ are shown in Figure \ref{fig:03}.

There are significant increases in the detection rates of the Galactic events for BURSTT compared with STARE2 and GReX. In the upper panel of Figure \ref{fig:03}, for a model with YMW16 ISM and a lower energy cut-off of $10^{27}$ erg for the power-law ($\alpha = -0.3$), the event detection rate for BURSTT is higher by a factor of approximately 37 than STARE2 and the events are seen over distances from the Sun in the galactic disk of up to 20 kpc. On the other hand, the increased discovery rates are still obvious by a factor of approximately 4 compared with GReX. In the lower panel where the power-law of $\alpha = -0.5$ is applied, the event detection rate for BURSTT is higher by a factor of approximately 195 than STARE2, while the increased discovery rates by a factor of 12 compared with GReX.

We also tested the $\alpha=-1.0$ cases for STARE2, GReX and BURSTT. The relative detection rates for STARE2 are much smaller than BURSTT, while the relative detection rates for BURSTT increase by a factor of 172 compared with GReX. According to the radiometer equation (equations \ref{eq:08}), S/N is proportional to flux density (which is also proportional to energy). Therefore, steeper LFs produce many FRBs whose corresponding S/N is smaller and these FRBs may not be detected since they do not pass the detection threshold ($\sn > 10$). When $\alpha=-1.0$ is assumed, the ratio of detection rates between BURSTT and GReX becomes 14.3 times as high compared with the result with $\alpha=-0.5$: 12 $\times$ 14.3 $\sim$ 172 times more detections with BURSTT compared with GReX. It is because the steeper LF increases the number of faint FRB-like events which may be difficult to be detected with STARE2 and GReX due to their lower sensitivities.

In addition to less brighter FRB-like events, steeper LFs produce a larger number of fainter FRB-like events, which can only be detected closer to the Sun, especially for the detection rates with STARE2 and GReX. Our results indicate that the relative improvements on the detection rates depend on the assumed power-law index. Notably, assuming $\alpha = -1.5$ is disfavoured because it is not consistent with the properties of the burst seen from SGR1935+2154 \citep{Bochenek2020b, Scholz2020}. Therefore, BURSTT can yield greater discovery rates of FRB-like events than both STARE2 and GReX within a reasonable range of assumed $\alpha$.

\section{Factors of specification of BURSTT in detection rate improvement}
\label{sec:Factors of specification of BURSTT in detection rate improvement}

In the previous section, we compare the number of FRB-like pulses detected by STARE2, GReX and BURSTT in our modelling and find that BURSTT has better detection rates than the other two surveys. In this section, we investigate how each telescope parameter of BURSTT can improve the detection rate of Galactic FRB-like events compared with STARE2. For this purpose, we change one of the system parameters of STARE2, such as the area of a single antenna, sensitivity, operating frequency, bandwidth and FoV, to that of BURSTT as a simulated system in our modelling. The detection rates under such assumptions are compared with that of STARE2.

\begin{figure*}
    \includegraphics[width=17cm]{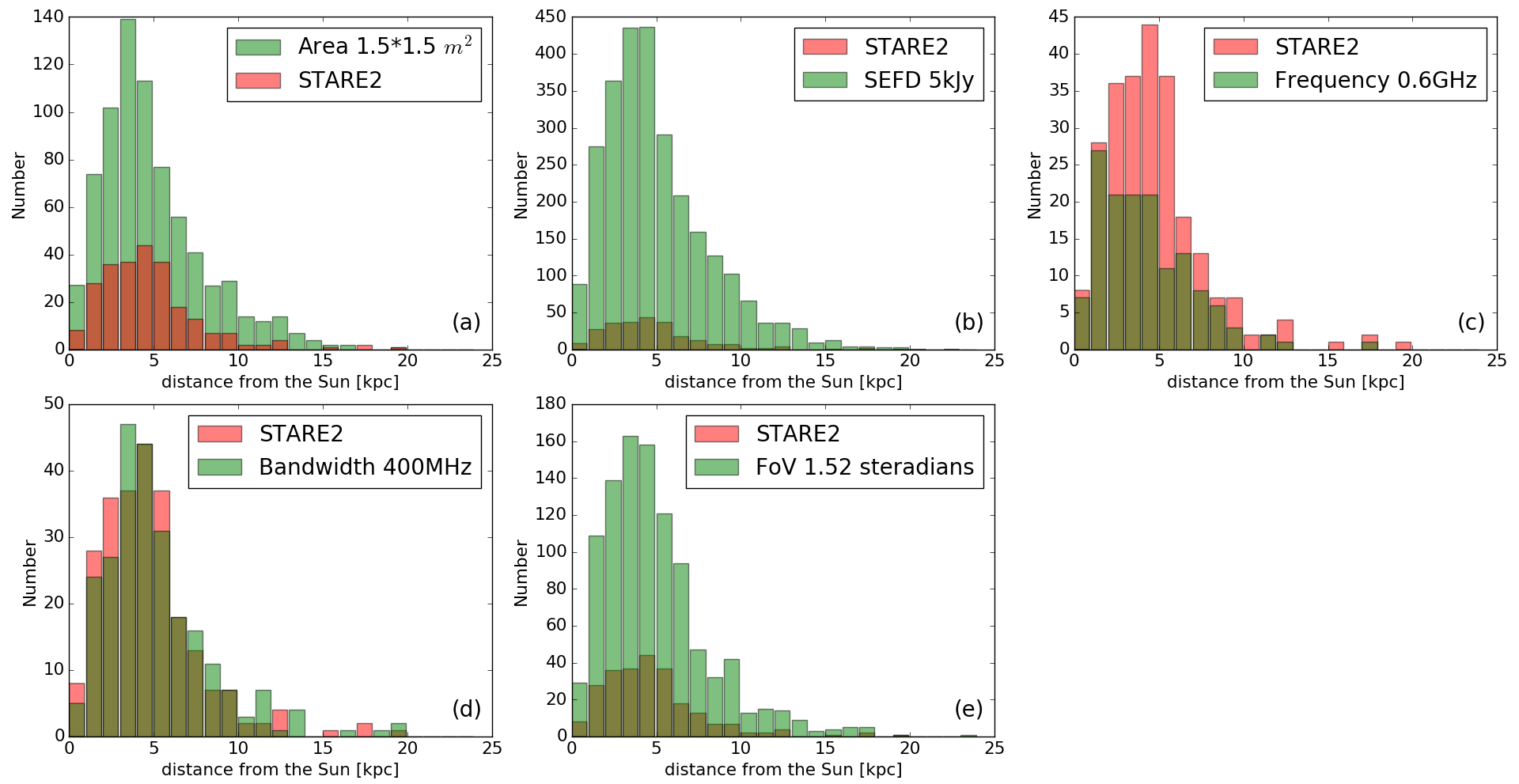}
    \caption{Comparison of detection rates between STARE2 and various simulated telescopes for Model II, and $E_0 = 10^{27}$ erg for the LF with $\alpha = -0.3$. The vertical axis represents the number of FRB-like events that pass the detection threshold $\sn > 10$ and width cut $\tau < 100$ ms.}
    \label{fig:04}
\end{figure*}

The results of a typical run for Model II with a power-law $\alpha = -0.3$ and low energy cut-off $E_0 = 10^{27}$ erg are shown in Figure \ref{fig:04}. In all panels, red represents the number of events predicted for STARE2 while green represents the detection rate of STARE2 with an improved parameter adopted from BURSTT. Area of a single antenna, SEFD, operating frequency, bandwidth and FoV are altered in panels (a) to (e) respectively.

We find that in panel (a) of Figure \ref{fig:04}, the simulated system that is applied with the value of BURSTT’s antenna area has an obvious increase in detection rate by a factor of 3 and the events are observed over distances from the Sun in the galactic disk of up to 23 kpc. In panel (b), for the system applied with the BURSTT’s SEFD, the increased discovery rates are significant by a factor of approximately 11 and the events that are located at 23 kpc away from the Sun are observable. However, in panel (c), the simulated system has a lower detection rate than STARE2 with a factor of 0.6. We note that the observational frequency of BURSTT is lower than that of STARE2. Therefore, scattering and smearing are more significant at the BURSTT's frequency, reducing S/N through their pulse-broadening effects \citep[][see also equation \ref{eq:04} and \ref{eq:06}]{Bhat2004}. In panel (d), there is only a modest difference for systems applied with the value of bandwidth of BURSTT, by a factor of 1. The detection rates are improved by a factor of 4 by adopting BURSTT’s FoV in panel (e). The summary of the factor of detection rate improvements by applying various parameters of BURSTT is shown in Table \ref{table:2}.

\begin{table*}
    \centering
    \begin{tabular}{lcr} 
     \hline
     \hline
     System parameters & Ratio of value compared with STARE2 & Ratio of detection rate \\
     \hline
     \hline
     Area of antenna & $\times$ 56 larger & 3 \\ 
     \hline
     SEFD & $\times$ 20000 more sensitive & 11 \\ 
     \hline
     Central frequency & $\times$ 2 smaller & 0.6 \\ 
     \hline
     Bandwidth & $\times$ 2 larger & 1 \\
     \hline
     Field of view (FoV) & $\times$ 1.4 larger & 4 \\
     \hline
     \hline
     \multicolumn{2}{|r|}{Expected detection rate} & 26.4 \\
     \hline
     \hline
    \end{tabular}
    \caption{Specifications of BURSTT used in comparing with STARE2 and their factor of discovery rate improvements. The first column shows the system parameters of STARE2 which is changed to that of BURSTT as a simulated system in our modelling, the second column shows the ratio of each corresponding parameter compared with that of STARE2 and the third column shows the ratio of detection rate of FRB-like events between corresponding simulated system and STARE2. The last row shows the expected detection rate of BURSTT calculated by $11 \times 0.6 \times 1 \times 4$ and the result is close to 37 mentioned in Section \ref{sec:Comparison of detection rates among STARE2, GReX and BURSTT}. The power-law slope of $\alpha=-0.3$ is assumed on the luminosity function.}
    \label{table:2}
\end{table*}

In conclusion, with the ISM model NE2001, YMW16 and the luminosity function adopted, the sensitivity of BURSTT has a significant effect on the improvement in the detection rate of FRB-like events in the Milky Way. In the future, it will become a primary consideration for proposing deeper, high-spatial-resolution galactic FRB-like events surveys.

\section{BURSTT with 2048 antennas}
\label{sec:BURSTT with 2048 antennas}

\begin{figure}
    \begin{subfigure}{10cm}
    \includegraphics[width=9.0cm]{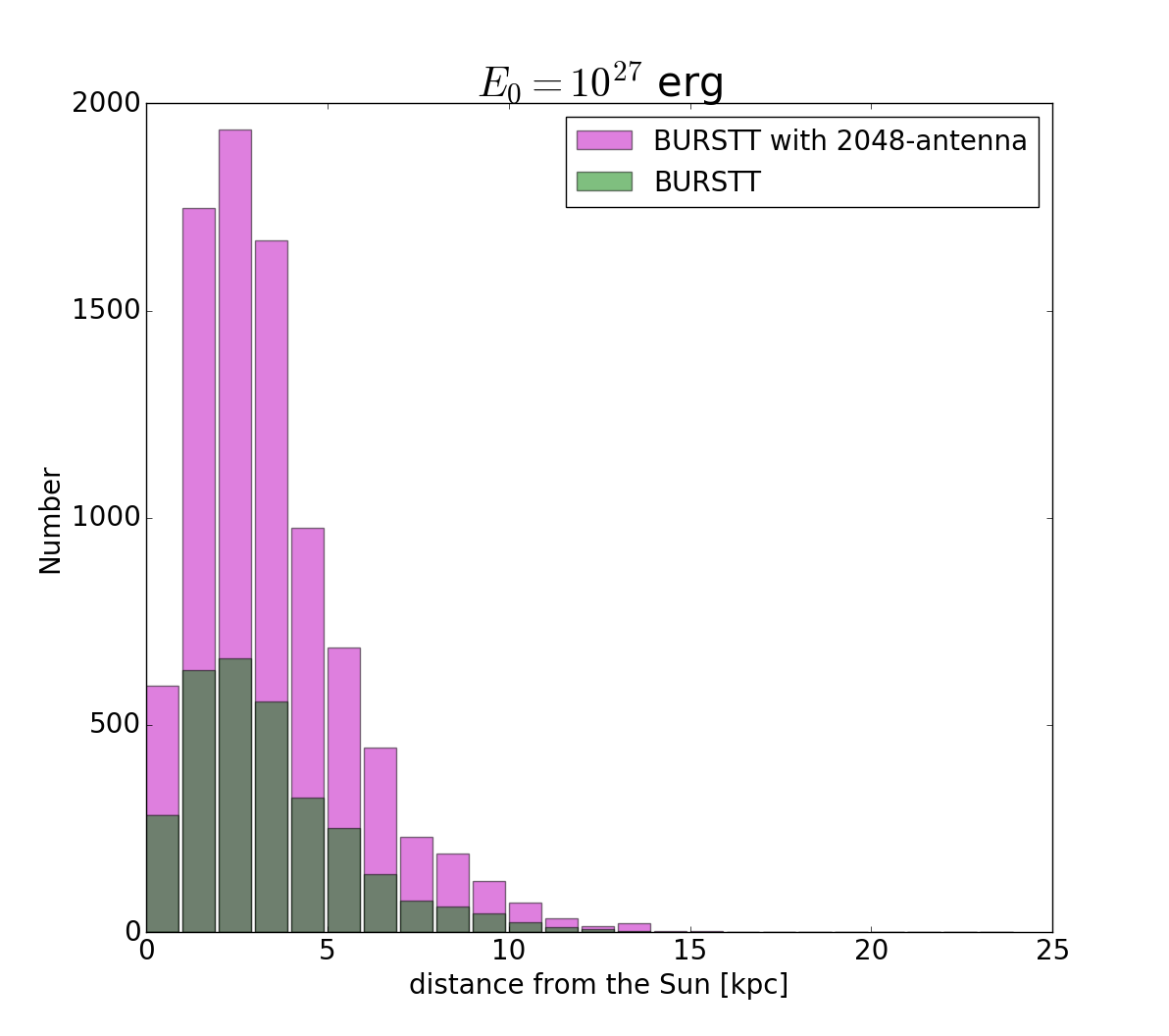}
    \end{subfigure}
    \begin{subfigure}{10cm}
    \includegraphics[width=9.0cm]{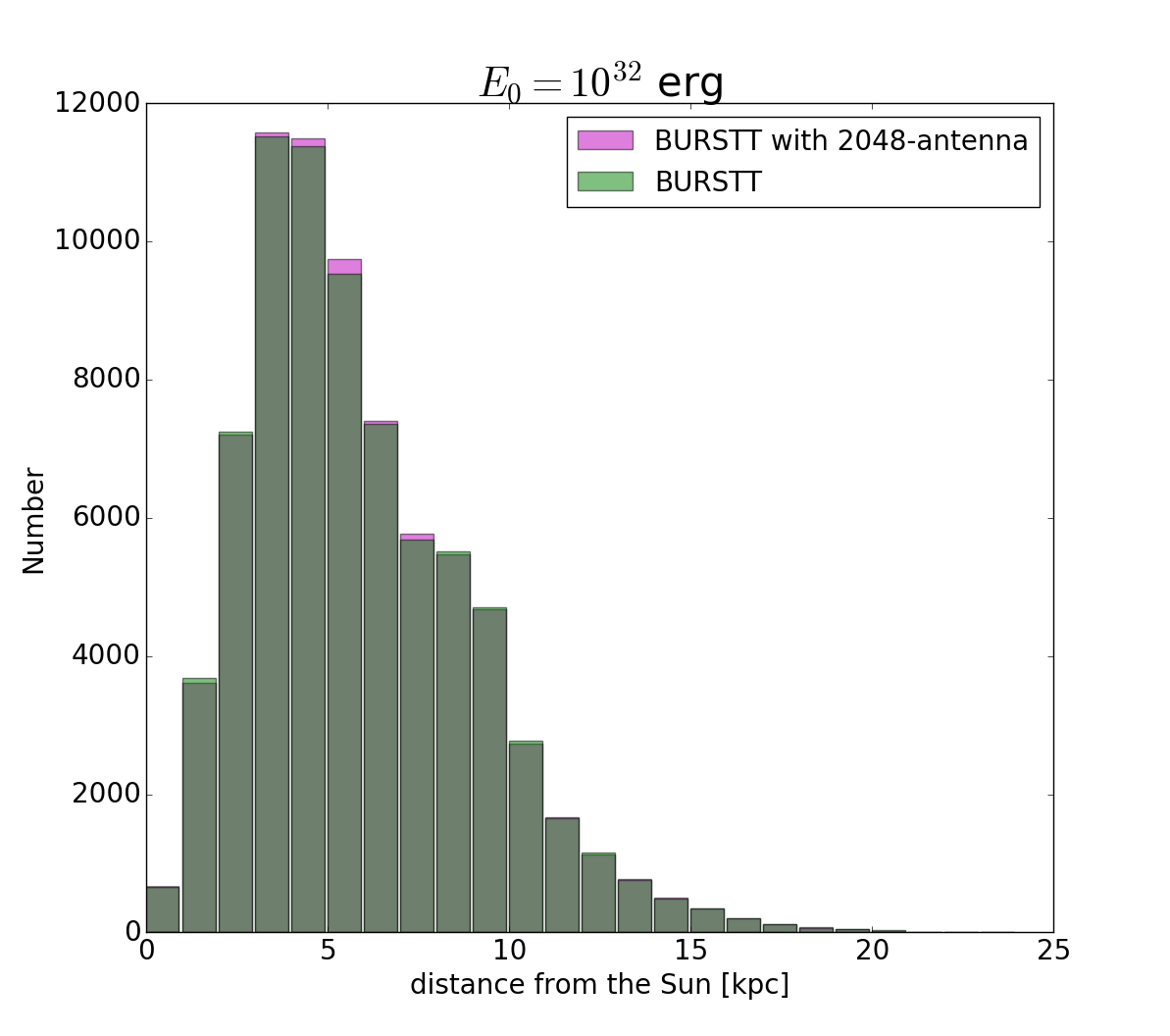}
    \end{subfigure}
    \caption{Upper panel: Comparison of BURSTT and BURSTT-2048 detection rates for Model II, $E_0 = 10^{27}$ erg for the LF with $\alpha = -0.5$. The vertical axis represents the number of FRB-like events that pass both the detection threshold ($\sn > 10$) and width cut ($\tau < 100$ ms). Lower panel: Same as the upper panel but with an LF with $E_0 = 10^{32}$ erg.}
    \label{fig:05}
\end{figure}

While writing this paper, the upgrade version of BURSTT, BURSTT-2048 consisting of 2048 antenna, is proposed to increase the FRB-like events rate per year \citep{Lin2022}. The larger number of antennas provides a sensitivity which is about 8 times better than the original BURSTT (which has SEFD of 5 kJy). Therefore, BURSTT-2048 has an SEFD of about 600 Jy. In this section, we examine the detection rate improvement of BURSTT-2048 by applying the proposed system parameters of BURSTT and BURSTT-2048 \citep{Lin2022} in our models with power-law slopes $\alpha= -0.5$. A range of energy cut-offs of $E_0 = 10^{27}$ and $10^{32}$ ergs is adopted in the simulation to examine its effect on the detection rates of FRB-like events. The results of simulations for Model II with energy cut-offs $E_0 = 10^{27}$ and $10^{32}$ ergs are shown in Figure \ref{fig:05}.

As we expected, there are significant increases in the detection rates of the low luminosity Galactic events for BURSTT-2048 compared with BURSTT. In the upper panel of Figure \ref{fig:05}, by applying a lower energy cut-off of $10^{27}$ erg, the event detection rate for BURSTT-2048 is higher by a factor of approximately 3 than BURSTT and the events are seen over distances from the Sun in the galactic disk of up to 15 kpc. In the lower panel where the energy cut-off of $10^{32}$ erg is applied, there is no significant difference between BURSTT and BURSTT-2048. These results indicate that the sensitivity of BURSTT is high enough to detect Galactic FRB-like events with $>10^{32}$ erg, whereas BURSTT-2048 will significantly increase samples fainter than $10^{32}$ erg.

\section{Conclusion}
\label{sec:Conclusion}

We have examined the prospects for searching Galactic FRB-like events in the Milky Way by generating a population of mock FRB-like event sources. We use the publically available code, \texttt{MilkyWay-FRBs} \citep{Gohar2022}, to model the spatial distribution of the FRB-like events in the Milky Way disk and the effects of the ISM in the detection of the events, such as broadening of the pulse widths and DM smearing. The \texttt{MilkyWay-FRBs} code introduces a power-law FRB-like event luminosity function into the models as well and simulates the BURSTT experiment by adopting the system parameters of BURSTT.

A comparison of detection rates of FRB-like events among STARE2, GReX and BURSTT is performed for a range of luminosity functions. As a result, BURSTT has the potential to increase the detection rates by more than two orders of magnitude compared with STARE2 and GReX, depending on the slope of the luminosity function of FRB-like events. A comparison of FRB-like events detection rates between BURSTT and BURSTT with 2048 antenna (BURSTT-2048) is simulated as well and BURSTT-2048 has a greater improvement in the detection rate of fainter events by a factor of 3. Such a significant increasement of the Galactic FRB-like sample would shed light on the general picture of the FRB origin.

We investigate the influence of the specifications of BURSTT on its detection improvement as well, by comparing the discovery rates among the simulated systems with various system parameters. We realise that BURSTT has the best ability to collect a large sample of nearby FRB-like events with accurate positions because of its greatly higher sensitivity and improved coverage of the Milky Way plane relative to STARE2 and GReX. Along with BURSTT being in function in the future, we expect to find multiple new FRB-like events from sources including Galactic magnetars each year.

\section*{Acknowledgements}

We thank the referee, Chris Flynn, for many insightful comments, which improved the paper significantly. The BURSTT project is supported by the National Science and Technology Council of Taiwan through Science Vanguard Research Program 111-2123-M-001-008-. We thank Hsiu-Hsien Lin and Ue-Li Pen for useful discussion on the BURSTT specification. TG acknowledges the supports of the National Science and Technology Council of Taiwan through grants 108-2628-M-007-004-MY3 and 111-2123-M-001-008-. TH acknowledges the supports of the National Science and Technology Council of Taiwan through grants 110-2112-M-005-013-MY3, 110-2112-M-007-034-, and 111-2123-M-001-008-.

\section*{Data Availability}
The data underlying this article is generated by the publically available code, \texttt{MilkyWay-FRBs}  \citep{Gohar2022}. The code is available at \url{https://github.com/cmlflynn/milkyway-frbs}.

\bibliographystyle{mnras}
\bibliography{bibliography}

\appendix

\section{Testing difference slopes of luminosity functions}
\label{sec:Testing difference slopes of luminosity functions}

\begin{figure*}
    \includegraphics[width=17cm]{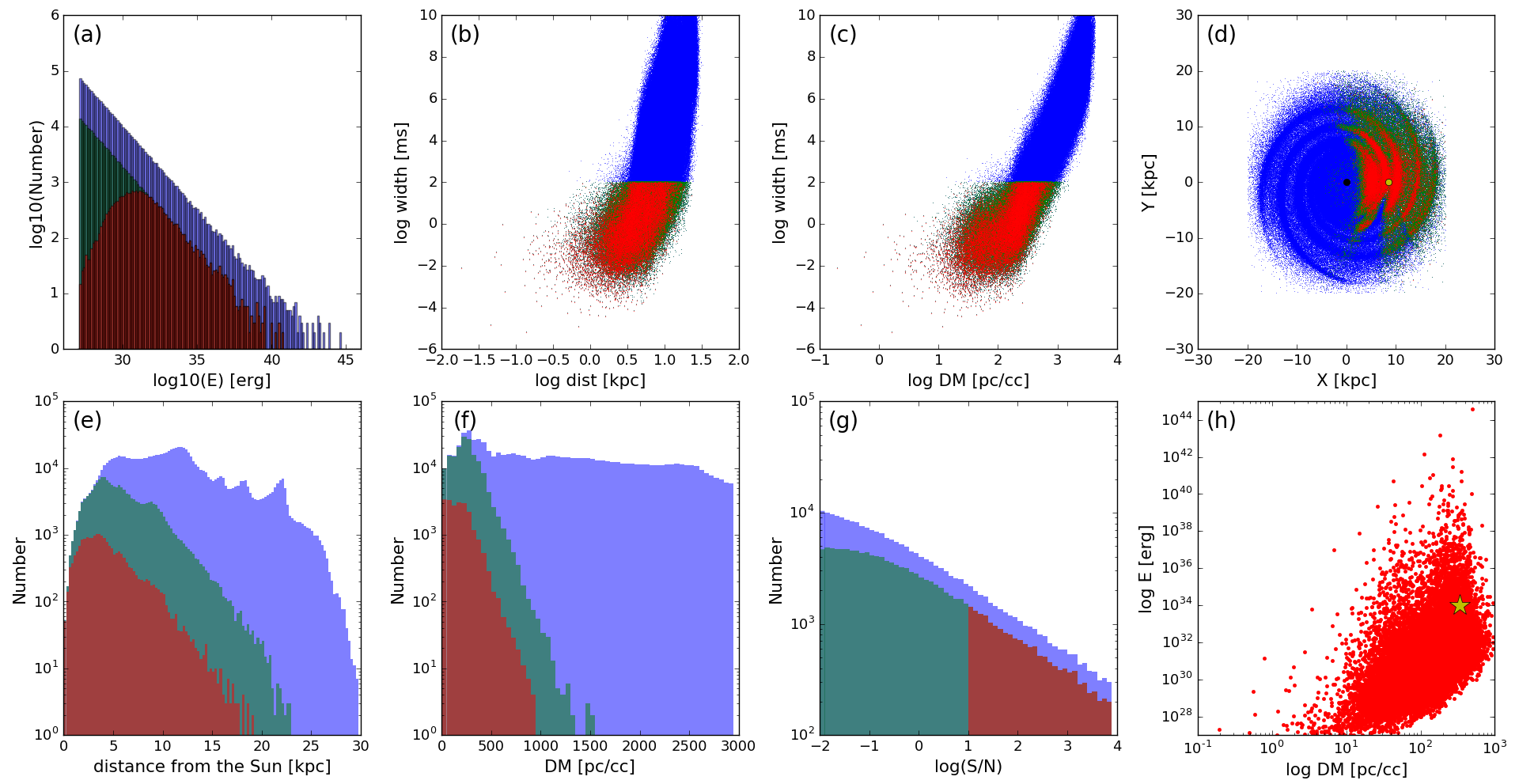}
    \caption{Simulated FRB-like events in the Milky Way for Model II, with power-law index $\alpha = -0.3$ and lower energy cut-off $E_0 = 10^{27}$ erg. In all panels, blue represents all FRB-like events simulated, green represents those that pass the width cut $\tau < 100$ ms and red represents those that pass the detection threshold $\sn > 10$ and $\tau < 100$ ms. The yellow and black dots in panel (d) represent the Sun and the Galactic center, respectively. The yellow star symbol in panel (h) represents the Galactic FRB event from SGR 1935+2154 detected in April 2021.}
    \label{fig:06}
\end{figure*}

\begin{figure*}
    \includegraphics[width=17cm]{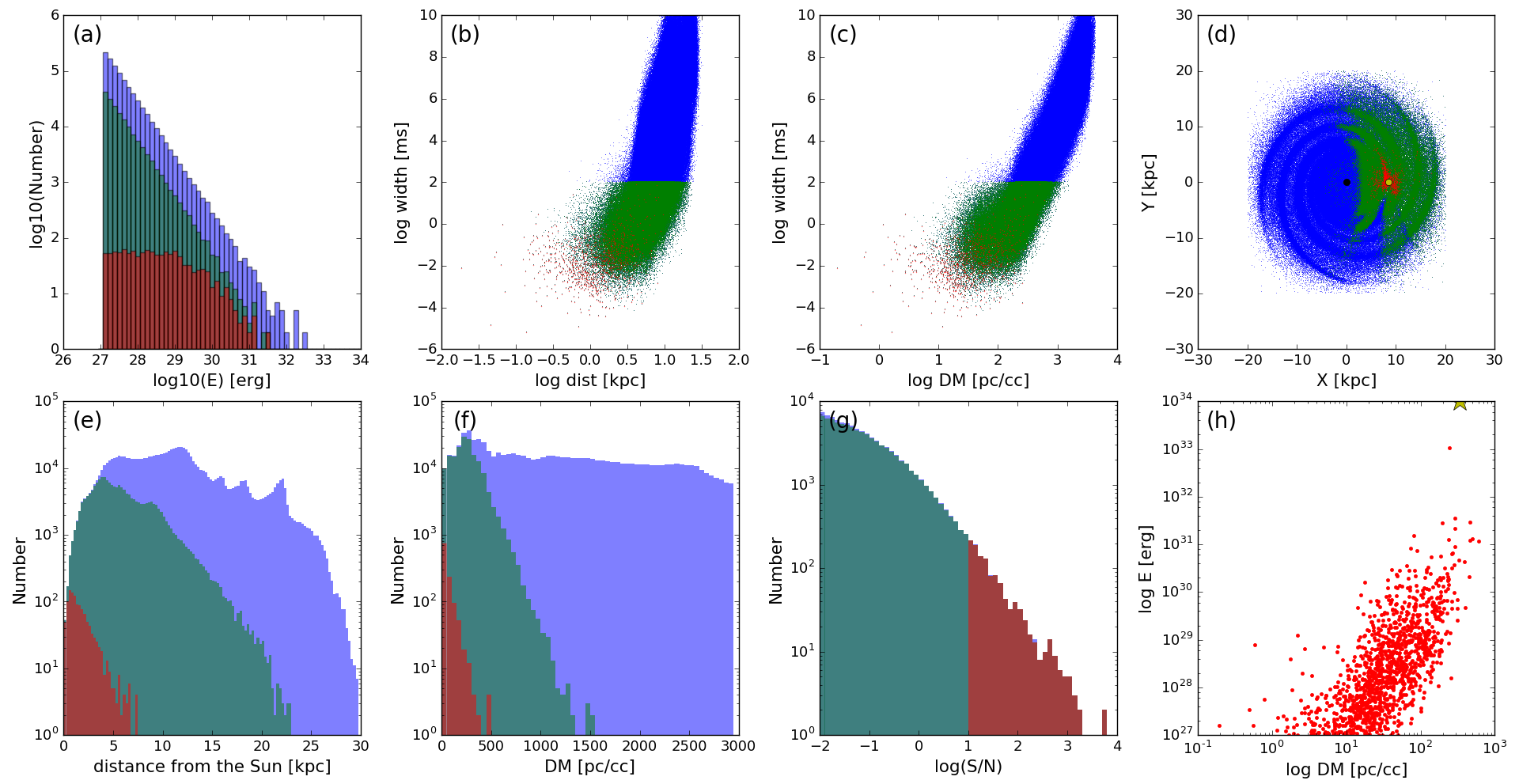}
    \caption{Simulated FRB-like events in the Milky Way for Model II, with power-law index $\alpha = -1.0$ and lower energy cut-off $E_0 = 10^{27}$ erg. All panels are the same as Figure \ref{fig:06}.}
    \label{fig:07}
\end{figure*}

In Section \ref{subsec:Simulation of mock FRB-like events in Milky Way}, we ran the simulations of mock FRB-like events for a lower energy cut-off $E_0$ of $10^{27}$  erg and a range of power-law slopes of $\alpha= -0.3, -0.5$ and $-1.0$ to examine whether sufficiently deep sampling of the luminosity function (LF) are probed. The results of a run of simulation of FRB-like events for Model II with a power-law of $\alpha = -0.3$ and $-1.0$ are shown in Figures \ref{fig:06} and \ref{fig:07} respectively.

Based on Figure \ref{fig:06}, a detectable FRB-like events by BURSTT has an energy of approximately $10^{27}$ to $10^{41}$ erg and a DM in the range of 0 to 900 $\rm{pc\,{cm}^{-3}}$. On the other hand, based on Figure \ref{fig:07}, a detectable FRB-like events has an energy in the range of approximately $10^{27}$ to $10^{31}$ and a DM of 0 to 500 $\rm{pc\,{cm}^{-3}}$. For both Figures \ref{fig:06} and \ref{fig:07}, the detectable FRBs are located in the range between 0 and 20 kpc from the Sun.

By comparing Figures \ref{fig:06} and \ref{fig:07} with Figure \ref{fig:02} (which is shown in Section \ref{subsec:Simulation of mock FRB-like events in Milky Way}), we find that in panel (a) of Figures \ref{fig:02} and \ref{fig:06}, the results of simulation with $\alpha = -0.5$ and $-0.3$ show the significant “turns over” distributions, that is low luminosity FRB-like events have started to decline again after a steep rise to a maximum, implying that an adequate number of intrinsically fainter but nearby detectible samples are generated. In panel (a) of Figure \ref{fig:07}, the result of simulation with $\alpha = -1.0$ is less significant, while the distributions include starting points of the turnover. Apart from this, all the results in panel (h) are consistent with the properties of the burst seen from SGR1935+2154. The results of simulations with $\alpha = -1.0$ (which is panel (h) in Figure \ref{fig:07}) are also reasonably close to the SGR1935+2154 event.

In summary, the simulations of FRB-like events in the Milky Way for Model II with $\alpha = -0.3$ and $\alpha = -0.5$ show better statistical results. We put the result of simulation with $\alpha = -0.5$ as a representative figure in Section \ref{subsec:Simulation of mock FRB-like events in Milky Way}.

\bsp
\label{lastpage}
\end{document}